\documentstyle{agile7th}

\input epsf.sty
\input epsfig.sty
\input psfig.sty

\def \AAP #1 #2 {{\em Astron. Astrophys.\/} {\bf #1}, #2}
\def \AAL #1 #2 {{\em Astron. Astrophys. Lett.\/} {\bf #1}, L#2}
\def \AAR #1 #2 {{\em Astron. Astrophys. Rev.\/} {\bf #1}, #2}
\def \AAS #1 #2 {{\em Astron. Astrophys. Suppl. Ser.\/} {\bf #1}, #2}
\def \AJ #1 #2 {{\em Astron. J.\/} {\bf #1}, #2}
\def \ANNREV #1 #2 {{\em Ann. Rev. Astron. Astrophys.\/} {\bf #1}, #2}
\def \APJ #1 #2 {{\em Astrophys. J.\/} {\bf #1}, #2}
\def \APJL #1 #2 {{\em Astrophys. J. Lett.\/} {\bf #1}, L#2}
\def \APJS #1 #2 {{\em Astrophys. J. Suppl.\/} {\bf #1}, #2}
\def \APSS #1 #2 {{\em Astrophys. Space Sci.\/} {\bf #1}, #2}
\def \ASR #1 #2 {{\em Adv. Space Res.\/} {\bf #1}, #2}
\def \BAIC #1 #2 {{\em Bull. Astron. Inst. Czechosl.\/} {\bf #1}, #2}
\def \JSQRT #1 #2 {{\em J. Quant. Spectrosc. Radiat. Transfer\/} {\bf #1}, #2}
\def \MN #1 #2 {{\em Mon. Not. R. Astr. Soc.\/} {\bf #1}, #2}
\def \MEM #1 #2 {{\em Mem. R. Astr. Soc.\/} {\bf #1}, #2}
\def \PLR #1 #2 {{\em Phys. Lett. Rev.\/} {\bf #1}, #2}
\def \PASJ #1 #2 {{\em Publ. Astron. Soc. Japan\/} {\bf #1}, #2}
\def \PASP #1 #2 {{\em Publ. Astr. Soc. Pacific\/} {\bf #1}, #2}
\def \NAT #1 #2 {{\em Nature\/} {\bf #1}, #2}
\def \SAIT #1 #2 {{\em Mem.\ Soc.\ Astron.\ It.\/} {\bf #1}, #2}
\def \MESS #1 #2 {{\em The Messenger\/} {\bf #1}, #2}
\def \ASTRNACH #1 #2 {{\em Astron. Nach.\/} {\bf #1}, #2}
\def \AGPSR #1 #2 {{\em ASI Special Publication\/} {\bf #1}, #2}

\def \PRL #1 #2 {{\em Physical Review Letters\/} {\bf #1}, #2}

\begin{opening}

\title{A search for VHE counterparts of galactic Fermi sources}
\author{P. H. T. Tam$^{1}$,  S. J. Wagner$^{1}$, O. Tibolla$^{2}$, R. C. G. Chaves$^{2}$}
\institute{$^1$Landessternwarte, Universit\"at Heidelberg, K\"onigstuhl, D 69117 Heidelberg, Germany\\
$^2$Max-Planck-Institut f\"ur Kernphysik, P.O. Box 103980, D 69029 Heidelberg, Germany}
\date{} % DO NOT INSERT ANY DATE HERE !!!
\end{opening}

\begin{document}

%\oddpagefooter{\sf $7^{th}$ AGILE Workshop}{}{\thepage}
%\evenpagefooter{\thepage}{}{\sf $7^{th}$ AGILE Workshop}
\oddpagefooter{}{}{} % LEAVE AS IT IS !
\evenpagefooter{}{}{} % LEAVE AS IT IS !
\medskip  % LEAVE AS IT IS !

\begin{abstract} % LEAVE THIS COMMAND AS IT IS AND WRITE THE ABSTRACT IN THE FOLLOWING!
Very high-energy (VHE; E$>$100~GeV) gamma-rays have been detected from a wide range of
astronomical objects, such as SNRs, pulsars and
pulsar wind nebulae, AGN, gamma-ray binaries, molecular clouds, and
possibly star-forming regions as well.
%, thanks to the high sensitivity of the current generation of Imaging Atmospheric Cherenkov Telescopes (IACTs).
At lower energies, sources detected using Large Area Telescope (LAT) aboard Fermi provide a rich set of data
which can be used to study the behavior of cosmic accelerators
in the GeV to TeV energy bands. In
particular, the improved angular resolution in both bands compared to previous instruments significantly
reduces source confusion and facilitates the identification of associated
counterparts at lower energies.
%This allows detailed studies of individual sources to be carried out.
In this proceeding, a comprehensive search for VHE gamma-ray sources which are
spatially coincident with Galactic Fermi/LAT bright sources is performed, and the GeV to TeV spectra of selected coincident sources are shown. It is found that LAT bright GeV sources are correlated to TeV sources, in contrast with previous studies using EGRET data. %Moreover, a single spectral component seems unable to describe the MeV to TeV spectra of some coincident GeV/TeV sources.
\end{abstract}

\medskip

\section{Introduction}

The knowledge of the very high-energy (VHE; E $>$100~GeV) sky has greatly improved during the last few years, thanks to the high sensitivity of current imaging atmospheric Cherenkov telescopes (IACTs), e.g. H.E.S.S., MAGIC, and VERITAS. At lower energies, observations using $\gamma$-ray satellites like \emph{CGRO}/EGRET, \emph{AGILE}, and \emph{Fermi}/LAT represent the dominant efforts in the field of high-energy $\gamma$-ray astronomy.

%Gamma-ray observations of Galactic sources can help us to solve a number of important astrophysical questions, including (1) the physics of pulsars, PWN and SNR; and (2) the origin of cosmic rays. To fully explore the It seems natural to carry out a comparison of sources detected in the $\sim$100~MeV--100~GeV and those in the $>$100~GeV range.

Funk et al. (2008) compared $\gamma$-ray sources in the third EGRET catalogue~\citeyear{3rd_egret_cat} and those 22 then-published H.E.S.S. sources within the region of $l=-$30$^\circ$ to 30$^\circ$, $b=-$3$^\circ$ to 3$^\circ$~\citeyear{hess_survey}. They did not find spatial correlation between the two populations. However, due to the capabilities of the EGRET experiment, this study suffers from the following limitations: (1) The sensitivity of EGRET is relatively poor. The lack of photon statistics leads to poor-constrained spectral indices and the spectra end $\sim$10~GeV at the upper end for a typical source; (2) EGRET sources are only localized to degree-scales, which is much larger than the angular resolution of IACTs. These disadvantages are now largely improved by the better performance of LAT over EGRET.

In February 2009, the \emph{Fermi}/LAT bright source list (BSL) was released~\citeyear{bsl_lat}. In this work, VHE counterparts detected using IACTs of each source in this list are searched for based on spatial coincidence, and the GeV to TeV spectra of several selected coincident sources are depicted.

\section{Search for spatial coincidence}

\subsection{The Fermi and VHE catalogues}

Abdo et al. (2009a) present 205 point-like $\gamma$-ray sources detected in the $0.2-100$~GeV band based on three months of observations (August~4, 2008 -- October~30, 2008). The authors assign for each source a source class, as well as $\gamma$-ray and lower energy association (if any). Those sources that are classified as extragalactic (all AGN and the Large Magellanic Cloud) are not considered below. Most of the LAT bright sources studied in this work are, therefore, of Galactic origin.

The remaining source list contains 83 sources, that consists of 15 radio/X-ray pulsars, 15 new pulsars discovered using LAT, 2 high-mass X-ray binaries (HMXBs), one globular cluster (47 Tuc), 13 SNR/PWN candidates, and 37 sources without obvious counterparts at lower energy bands (Unids) \cite{bsl_lat}.

The number of VHE $\gamma$-ray sources is larger than 50~\cite{VHE_Review_08,hess_survey08,hess_survey09}. The VHE $\gamma$-ray source positions and source extension in this work are taken from the corresponding publications. At higher energies, the Milagro collaboration reported evidence of multi-TeV emission from several LAT source positions~\cite{milagro_bsl}. Only those source candidates with significance larger than five are regarded as TeV sources here. With several tens of known sources in both the GeV and TeV bands, a systematic cross-correlation study is conducted.

% published H.E.S.S.\footnote{An online H.E.S.S. source catalogue can be found at http://www.mpi-hd.mpg.de/hfm/HESS/pages/home/sources/.}

\subsection{Level of spatial coincidence}
\label{position_procedure}

To quantify the level of spatial coincidence, the following scheme is employed. Let $d$ be the distance between a LAT best-fit centroid position and a nearby VHE source centroid position. The radius of 95\% confidence region for the LAT source is the uncertainty on the centroid position of the given LAT source, which is typically $\sim0.1^\circ$. Most VHE sources are extended, with a typical extension of $0.1^\circ-0.5^\circ$. Let $e$ be the sum of the radius of 95\% confidence region and the source extension of the VHE source.

For each LAT source, if a VHE source was found where $d - e < 0$, the source pair is called a spatial coincidence case (i.e. category \emph{Y} -- Yes). Given a possible extended nature of many LAT bright sources, so that the sources seen by LAT and VHE instrument may actually overlap with each other, a category \emph{P} (for Possible) is defined for pairs where $0 < d - e < 0.3^\circ$. If no reported VHE source was found with $d - e < 0.3^\circ$: the LAT source falls into the coincidence level \emph{N} (for No coincidence with any VHE source). If there are several nearby VHE sources, only the closest VHE source would be considered.

\subsection{Spatial coincidence GeV/TeV pairs}
\label{position_results}

In the search, 23 coincident cases (Y, among them three are coincident with Milagro source only) and 5 possibly-coincident cases (P) are found. The results are presented in Tables~\ref{tab:pos_coincidence1}, \ref{tab:pos_coincidence2}, and \ref{tab:pos_coincidence3}. No reported VHE sources are found in the remaining 55 sources, as of October 2009.
%%%%%%%%%%%%%%%%%%%%%%%%%%%%%%%%%%%%%%%%%% Table 1
\begin{table}[ht]
\caption{0FGL sources with spatially coincident VHE counterpart. PSR: pulsars; SNR/PWN: supernova remnants/pulsar wind nebulae candidates; Unid: unidentified sources; HMXB: high-mass X-ray binaries. }
\label{tab:pos_coincidence1}
\begin{center}
\begin{tabular}{lcc|l}
\hline\hline
     LAT source        & association & class & VHE source \\
    \hline
    0FGL~J0534.6+2201 & Crab    & PSR & HESS J0534+220  \\
    0FGL~J0835.4--4510 & Vela    & PSR & HESS J0835--455  \\
    0FGL~J1418.8--6058 &         & PSR & HESS J1418--609  \\
    0FGL~J1709.7--4428 & PSR B1706--44 & PSR & HESS J1708--443 \\
    0FGL~J1907.5+0602 &         & PSR & HESS J1908+063  \\
    0FGL~J2032.2+4122 &         & PSR & TeV J2032+4130  \\
    0FGL~J0617.4+2234 &         & SNR/PWN & VER J0616.9+2230  \\
    0FGL~J1615.6--5049 &         & SNR/PWN & HESS J1616--508  \\
    0FGL~J1648.1--4606 &         & SNR/PWN & Westerlund 1 region  \\
    0FGL~J1714.7--3827 &         & SNR/PWN & HESS J1714--385  \\
    0FGL~J1801.6--2327 &         & SNR/PWN & HESS J1801--233  \\
    0FGL~J1834.4--0841 &         & SNR/PWN & HESS J1834--087  \\
    0FGL~J1923.0+1411 & W~51C    & SNR  & HESS J1923+141 \\
    0FGL~J1024.0--5754 &         & Unid & HESS J1023--575  \\
    0FGL~J1805.3--2138 &         & Unid & HESS J1804--216  \\
    0FGL~J1839.0--0549 &         & Unid & HESS J1841--055  \\
    0FGL~J1844.1--0335 &         & Unid & HESS J1843--033  \\
    0FGL~J1848.6--0138 &         & Unid & HESS J1848--018  \\
    0FGL~J0240.3+6113 & LS I +61 303 & HMXB & VER J0240+612  \\
    0FGL~J1826.3--1451 & LS 5039      & HMXB & HESS J1826--148 \\
    \hline
\end{tabular}
\end{center}
\end{table}

%%%%%%%%%%%%%%%%%%%%%%%%%%%%%%%%%%%%%% Table 2
\begin{table}[ht]
\caption{0FGL sources with coincident MILAGRO source only }
\label{tab:pos_coincidence2}
\begin{center}
\begin{tabular}{lc|l}
\hline\hline
     LAT source        & class & Milagro source \\
    \hline
    0FGL~J0634.0+1745 & PSR & MGRO~C3   \\
    0FGL~J2020.8+3649 & PSR & MGRO~J2019+37 \\
    0FGL~J2229.0+6114 & PSR & MGRO~C4  \\
    \hline
\end{tabular}
\end{center}
\end{table}

%%%%%%%%%%%%%%%%%%%%%%%%%%%%%%%%%%%%%%%% Table 3
\begin{table}[ht]
\caption{0FGL sources with possibly coincident VHE source }
\label{tab:pos_coincidence3}
\begin{center}
\begin{tabular}{lc|l}
\hline\hline
     LAT source        & class & VHE source \\
    \hline
    0FGL~J1814.3--1739 & SNR/PWN & HESS J1813--178  \\
    0FGL~J1634.9--4737 & Unid & HESS J1634--472  \\
    0FGL~J1741.4--3046 & Unid & HESS J1741--302  \\
    0FGL~J1746.0--2900 & Unid & HESS J1745--290  \\
    0FGL~J1836.1--0727 & Unid & HESS J1837--069 \\
    \hline
\end{tabular}
\end{center}
\end{table}

The results are summarized as follows:
\begin{enumerate}
  \item Six LAT pulsars are spatially coincident with a source detected using IACTs, which may be the VHE-emitting PWN. In addition, three other have a MILAGRO counterpart, and have not been reported as detected using any IACTs yet.
  \item Among the 13 SNR/PWN candidates in the Fermi BSL, more than a half (7) are spatially-coincident with a VHE source, and additionally one being a possibly coincident case. The seemingly high fraction of coincidence is partly due to a better coverage of the inner Galaxy region with IACTs, where most SNR/PWN candidates are located. %This results in a generally better sensitivity of this class of sources than other classes.
  \item The two high-mass X-ray binaries listed in the BSL (0FGL~J0240.3+6113/LS~I +61~303 and 0FGL~J1826.3--1451/LS~5039) are both found spatially coincident with a VHE gamma-ray source, identified with the same corresponding binary.
  \item Five out of the 37 unidentified 0FGL~sources are spatially coincident with a VHE gamma-ray source. The number increases to nine if possible coincidence cases are included.
\end{enumerate}

% With such a large number of coincident cases, the relation of the GeV and TeV sources are explored. In the next section, the gamma-ray spectral energy distributions are constructed for those coincident and possibly coincident GeV/TeV source pairs with published VHE spectrum.

\section{The gamma-ray spectral energy distributions}

%\subsection{Construction of power law spectrum in the LAT range}
%\label{Sect:SED_construction}

Assuming a single power law, $F_{23}$ and $F_{35}$, photon flux in the low energy (10$^2$--10$^3$~MeV) and high energy (10$^3$--10$^5$~MeV) band, respectively \cite{bsl_lat}, are given by $F_{23}  = k \int_{0.1}^1 E^{-\Gamma} dE$ and $F_{35}  = k \int_1^{100} E^{-\Gamma} dE$, where $E$ is measured in GeV, $\Gamma$ is the photon index, and $k$ is the normalization at 1~GeV. From these two expressions, $k$ and $\Gamma$ can be obtained. The spectra in form of ``bowties'' are then constructed. %For those with $F_{23}$ given as a 2-$\sigma$ upper limit, the calculated $\Gamma$ can be treated as an upper limit and the reconstructed spectra can be seen as the ``softest possible'' power-law spectra. 
The power-law spectra are drawn from 100~MeV up to a certain maximum energy, $E_{\rm max}$ ($<$100~GeV), which is defined by requiring that the photon spectrum above $E_{\rm max}$ contains 10 photons over the three months of observations.%\footnote{Using the LAT on-axis effective area above 1~GeV of $\sim$8000~cm$^2$ and a mean on-axis exposure of $\sim$1~Ms~\cite{bsl_lat}}.
%$E_{\rm max}$ ranges from $\sim$3~GeV to 100~GeV.

Examples of the MeV--GeV spectra of spatially coincident GeV/TeV sources are depicted in Figures 1 to 3. Figure 1 shows four EGRET pulsars and their TeV-emitting nebula. Figure 2 shows the MeV to TeV spectra of two coincident GeV/TeV sources (0FGL~J1839.0-0549/HESS~J1841-055 and 0FGL~J1848.6-0138/HESS~J1848-018) in which the $\gamma$-ray spectra may be described by a single spectral component. Figure 3 shows the MeV to TeV spectra of two coincident GeV/TeV sources (0FGL~J1805.3-2138/HESS~J1804-216 and 0FGL~J1834.4-0841/HESS~J1834-087) in which two spectral components may be needed to describe the $\gamma$-ray spectra. It should be stressed that this notion is useful only if power-law spectra are good descriptions of the 0.1--$\sim$100~GeV source spectra.

\begin{figure}[ht!]
\centerline{\epsfig{figure=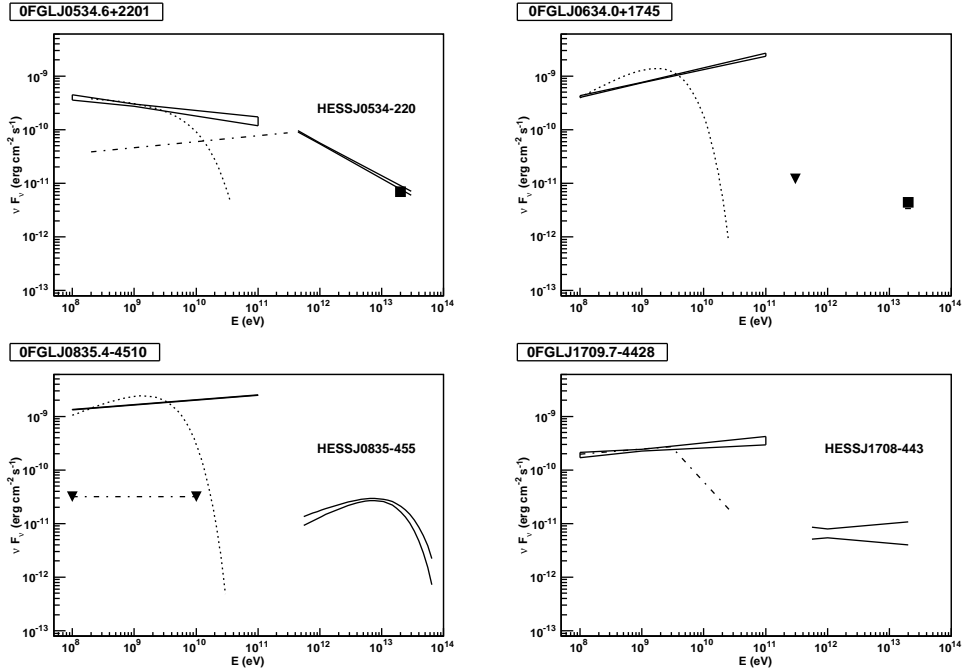,width=13.5cm,angle=0}}
\caption{Spectra of four EGRET pulsars and their TeV-emitting nebula. The solid lines at GeV energies represent the power-law representation described in Sect. 3.1. The ``cut-off'' GeV spectra up to several tens of GeV are the pulsed components, while the dashed-dotted lines in Crab (0FGL~J0534.6+2201) and Vela (0FGL~J0835.4--4510) are the nebula components (or upper limits thereof), both taken from the literatures.}

%\emph{Upper left}: Crab (0FGL~J0534.6+2201). The pulsar (dotted line) and nebula (dashed-dotted line) spectral components are those reported in~\cite{lat_crab}. The VHE spectra are taken from~\cite{hess_crab}, and the MILAGRO measurement at 20~TeV is shown~\cite{milagro_GPS}. \emph{Upper right}: Geminga (0FGL~J0634.0+1745). The pulsar spectrum (dotted line) are that reported in~\cite{lat_geminga}. The triangle denotes the upper limit reported in~\cite{veritas_geminga}, and the MILAGRO measurement at 20~TeV is also indicated~\cite{milagro_GPS}. \emph{Lower left}: Vela (0FGL~J0835.4--4510). The dotted line represents the Vela spectrum as shown in~\cite{lat_velapsr}, while the nebula component is constrained by the two triangles joined by the dashed-dotted line. The curved VHE spectrum is taken from~\cite{hess_velaX}. \emph{Lower right}: PSR~B1706--44 (0FGL~J1709.7--4428). The dashed-dotted line denotes the two power-law model spectrum derived in \cite{lat_egr_psr}. Both LAT energy spectra (though different above 3~GeV) are consistent with the photon flux in the 1--100~GeV band of this source~\cite{bsl_lat}. The VHE spectrum is taken from~\cite{hess_1706}.}
\end{figure}

\begin{figure}[ht!]
\centerline{\epsfig{figure=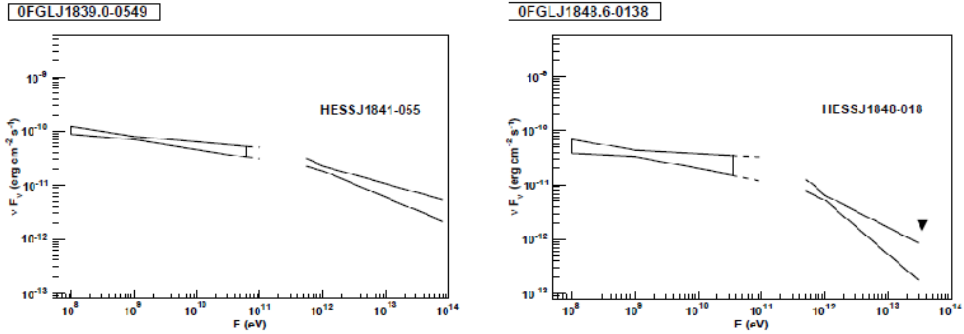,width=13.cm,angle=0}}
\caption{MeV to TeV spectra of two coincident GeV/TeV sources in which the $\gamma$-ray spectra may be described by a single spectral component }
\end{figure}

\begin{figure}[ht!]
\centerline{\epsfig{figure=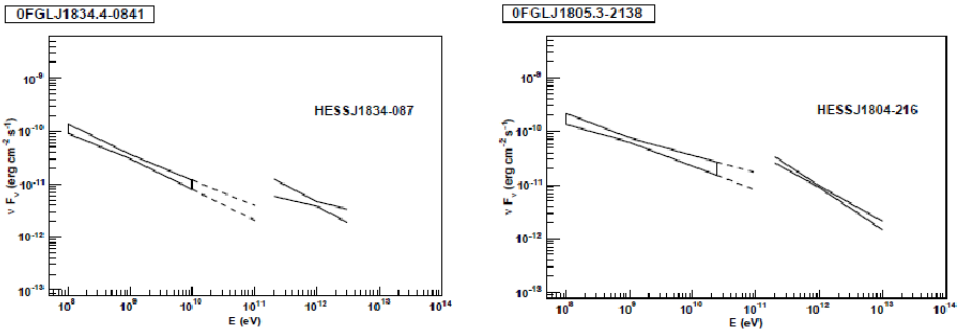,width=13.cm,angle=0}}
\caption{MeV to TeV spectra of two coincident GeV/TeV sources in which two spectral components may be needed to describe the $\gamma$-ray spectra }
\end{figure}

\section{The inner Galaxy region}

Although VHE observations only cover a small part of the whole sky, they do cover the majority of the inner Galaxy region. In particular, the H.E.S.S. telescopes have surveyed the region of $l=-$85$^\circ$ to 60$^\circ$, $b=-$3$^\circ$ to 3$^\circ$ up to summer 2009~\cite{hess_survey09}. %The fact that VHE counterparts are found at a higher fraction among LAT SNR/PWN candidates (7/13), as compared to pulsars (6/30), may indicate that not all LAT SNR/PWN candidates can be attributed to emission from unresolved gamma-ray pulsars. Otherwise, one expects a similar fraction from both populations. \emph{better quantify this assessment?}
In this region, there are 41 \emph{Fermi} bright sources. Among them, 16 are found coincident with a VHE counterpart. This fraction ($\sim$2/5) is higher than that for EGRET where about 1/4 of the EGRET sources (in a smaller region) are found to have a coincident VHE counterpart~\cite{funk08}. Moreover, the number raises to 21 if possibly coincident cases are included. LAT radii of 95\% confidence region are mostly much smaller than EGRET error boxes, which further strengthens the case. A breakdown of the number of coincidence cases for each source population in the region of $l=-$85$^\circ$ to 60$^\circ$, $b=-$3$^\circ$ to 3$^\circ$ is shown in Table~\ref{breakdown}.

%%%%%%%%%%%%%%%%%%%%%%%%%%%%%%%%%%%%%% Table 4
\begin{table}[ht]
\caption{Number of coincidence cases for each source population excluding AGN in the region $l=-$85$^\circ$ to 60$^\circ$, $b=-$3$^\circ$ to 3$^\circ$. The numbers in brackets include possibly coincident cases (P). }
\label{breakdown}
\begin{center}
\begin{tabular}{lcc}
\hline\hline
                &              & spatially    \\
LAT Source class  & 0FGL sources & coincident cases \\
\hline
pulsars              & 10      &  4      \\
SNR/PWN candidates   & 11      &  6 (7)   \\
Unidentified sources & 19      &  5 (9)   \\
\hline
Total (including LS~5039) & 41      &  16 (21) \\
\hline
\end{tabular}
\end{center}
\end{table}
%
%Use the following minipage example for the inclusion of tables.

%\vspace{.4cm} %TO ALLOW SUFFICIENT SPACE BETWEEN THE TEXT AND THE FIGURES
%
%\noindent
%\begin{minipage}{12.5cm}
%\centerline{\bf Tab. 2 - New possible identifications}
%\vspace{.5cm} %TO ALLOW SUFFICIENT SPACE BETWEEN THE TEXT AND THE FIGURES
%%
%
%\begin{tabular}{|c|c|c|c|c|c|c|c|}
%\hline
%
%Source & ID & CLASS & z & F$_{\gamma}^b$ & F$_{1keV}$ & F$_{5GHz}$ & \multicolumn{1}{|c|}{$\alpha_r$} \\
%
% Name$^a$& & & &$(\mu$Jy)  & $(\mu$Jy)   & (Jy)    & \\
%\hline
%
%0119+0312 & 0115+027 & QSO/OVV & 0.67  & 1.06 & 0.21 & 0.56$^c$
%& $-$0.77 \\
%0216+1107 & 0214+108 & QSO/OVV & 0.41 & 0.89 & 0.87 & 0.46$^c$
%& $-$0.73 \\
%0406+1704 & 0404+177 & QSO & 1.71 & 1.43 &  & 0.23$^c$
%& $-$0.32 \\
%0422+1414 & 0419+140 & & & 0.92 & & 0.40
%& $-$0.79  \\
%0432+2910 & 0430+2859 & OVV & & 1.19 & 0.23 & 0.48
%& \multicolumn{1}{|r|}{0.17} \\
%0809+5117 & 0806+524 & BL Lac & & 0.72 & 1.97 & 0.18
%& \multicolumn{1}{|r|}{0.07} \\
%\hline
%\end{tabular}
%
%\smallskip
%
%$^a$ IAU name; $^b$units of 10$^{-5} \mu$Jy; $^c$ Radio Loud = $P_{5GHz} > 5\times 10^{24}$ watt Hz$^{-1}$ \\
%
%\end{minipage}

\section{Conclusion}

This study, that largely benefits from the better angular resolution and better sensitivity of LAT over its predecessors including EGRET, for the first time, reveals a correlation between the GeV/TeV populations. The high fraction of \emph{Fermi} bright sources spatially coincident with a VHE counterpart cannot be a chance coincidence. This means that there exists a common GeV/TeV source population. On the other hand, a single spectral component is unable to describe some sources detected in both GeV and TeV energies. Two spectral components may be needed in these sources to accommodate the spectral energy distribution in $\gamma$-rays.


\begin{thebibliography}{}

\bibitem[Abdo et al.(2009a)]{bsl_lat} Abdo, A. A., Ackermann, M., Ajello, M., et al. (Fermi/LAT Collaboration), 2009a, \APJS 183 46
%\bibitem[Abdo et al.(2009b)]{latpsr_blind} Abdo, A. A., Ackermann, M., Ajello, M., et al. (Fermi/LAT Collaboration), 2009b, Science, 325, 840
%\bibitem[Abdo et al.(2009c)]{lat_velapsr} Abdo, A. A., Ackermann, M., Atwood, W. B., et al. (Fermi/LAT Collaboration), 2009c, \APJ 696 1084
%\bibitem[Abdo et al.(2009d)]{lat_LSI61} Abdo, A. A., Ackermann, M., Ajello, M., et al. (Fermi/LAT Collaboration), 2009d, \APJL 701 123
%\bibitem[Abdo et al.(2009e)]{lat_W51C} Abdo, A. A., Ackermann, M., Ajello, M., et al. (Fermi/LAT Collaboration), 2009e, \APJ, 706, 1
%\bibitem[Abdo et al.(2009f)]{lat_ls5039} Abdo, A. A., Ackermann, M., Ajello, M., et al. (Fermi/LAT Collaboration), 2009f, \APJL 706 56
\bibitem[Abdo et al.(2009b)]{milagro_bsl} Abdo, A. A., Allen, B.~T., Aune, T., et al. (MILAGRO Collaboration), 2009b, \APJL 700 127
%\bibitem[Abdo et al.(2007)]{milagro_GPS} Abdo, A.~A., Allen, B., Berley, D., et al. (MILAGRO Collaboration),\ 2007, \APJL, 664, L91
%%\bibitem[Abdo et al.(2007)]{milagro_cygnus_region} Abdo, A.~A., Allen, B., Berley, D., et al. (MILAGRO Collaboration), 2007, \APJL, 658, L33
%\bibitem[Acciari et al.(2009a)]{veritas_LSI61} Acciari, V.~A., Aliu, E., Arlen, T., et al. (VERITAS Collaboration)\ 2009a, \APJ, 700, 1034
%\bibitem[Acciari et al.(2009b)]{veritas_ic443} Acciari, V.~A., Aliu, E., Arlen, T., et al. (VERITAS Collaboration)\ 2009b, \APJL, 698, L133
%\bibitem[Aharonian et al.(2005a)]{aha05} Aharonian, F. A., Akhperjanian, A.~G., Aye, K.-M., et al. (H.E.S.S. Collaboration) 2005a, \AAL, 436, L17
%\bibitem[Aharonian et al.(2005b)]{hegra_TeV2032} Aharonian, F. A., Akhperjanian, A.~G., Beilicke, M., et al. (HEGRA Collaboration) 2005b, \AAP, 431, 197
%\bibitem[Aharonian et al.(2006a)]{aha06a} Aharonian, F. A., Akhperjanian, A.~G., Bazer-Bachi, A.~R., et al. (H.E.S.S. Collaboration) 2006a, \AAL, 448, L19
%\bibitem[Aharonian et al.(2006b)]{hess_velaX} Aharonian, F. A., Akhperjanian, A.~G., Bazer-Bachi, A.~R., et al. (H.E.S.S. Collaboration) 2006b, \AAL, 448, L43
%\bibitem[Aharonian et al.(2006c)]{hess_EBL_nature} Aharonian, F. A., Akhperjanian, A.~G., Bazer-Bachi, A.~R., et al. (H.E.S.S. Collaboration) 2006c, Nature, 440, 1018
%\bibitem[Aharonian et al.(2006d)]{hess_crab} Aharonian, F. A., Akhperjanian, A.~G., Bazer-Bachi, A.~R., et al. (H.E.S.S. Collaboration) 2006d, \AAP, 457, 899
%\bibitem[Aharonian et al.(2006e)]{hess_LS5039} Aharonian, F. A., Akhperjanian, A.~G., Bazer-Bachi, A.~R., et al. (H.E.S.S. Collaboration) 2006e, \AAP, 460, 743
\bibitem[Aharonian et al.(2006)]{hess_survey} Aharonian, F. A., Akhperjanian, A.~G., Bazer-Bachi, A.~R., et al. (H.E.S.S. Collaboration) 2006, \APJ 636 777
%\bibitem[Aharonian et al.(2006g)]{hess_kookaburra} Aharonian, F. A., Akhperjanian, A.~G., Bazer-Bachi, A.~R., et al. (H.E.S.S. Collaboration)\ 2006g, \AAP, 456, 245
%\bibitem[Aharonian et al.(2006h)]{hess_GC_DarkMatter} Aharonian, F. A., Akhperjanian, A.~G., Bazer-Bachi, A.~R., et al. (H.E.S.S. Collaboration)\ 2006h, \PRL, 97, 221102
%\bibitem[Aharonian et al.(2007a)]{hess_westerlund2_1023} Aharonian, F. A., Akhperjanian, A.~G., Bazer-Bachi, A.~R., et al. (H.E.S.S. Collaboration)\ 2007a, \AAP, 467, 1075
%\bibitem[Aharonian et al.(2007b)]{hess_psr_1718_1809} Aharonian, F. A., Akhperjanian, A.~G., Bazer-Bachi, A.~R., et al. (H.E.S.S. Collaboration)\ 2007b, \AAP, 472, 489
%\bibitem[Aharonian et al.(2008a)]{aha08} Aharonian, F. A., Akhperjanian, A.~G., Barres de Almeida, U., et al. (H.E.S.S. Collaboration) 2008a, \AAP, 477, 481
%\bibitem[Aharonian et al.(2008b)]{CTB_37A} Aharonian, F. A., Akhperjanian, A.~G., Barres de Almeida, U., et al. (H.E.S.S. Collaboration) 2008b, \AAP, 490, 685
%\bibitem[Aharonian et al.(2008c)]{hess_1801-233} Aharonian, F. A., Akhperjanian, A.~G., Barres de Almeida, U., et al. (H.E.S.S. Collaboration) 2008c, \AAP, 481, 401
%\bibitem[Aharonian et al.(2008d)]{hess_dark} Aharonian, F. A., Akhperjanian, A.~G., Barres de Almeida, U., et al. (H.E.S.S. Collaboration) 2008d, \AAP, 477, 353
\bibitem[Aharonian et al.(2008e)]{VHE_Review_08} Aharonian, F., Buckley, J., Kifune, T., \& Sinnis, G.\ 2008e, Reports on Progress in Physics \textbf{71}, 096901
%\bibitem[Aharonian et al.(2009)]{hess_1908+063} Aharonian, F. A., Akhperjanian, A.~G., Anton, G., et al. (H.E.S.S. Collaboration) 2009, \AAP, 499, 723
%\bibitem[Albert et al.(2007)]{magic_ic443} Albert, J., Aliu, E., Anderhub, H., et al. (MAGIC Collaboration) 2007, \APJL, 664, L87
%\bibitem[Albert et al.(2008)]{magic_crab_nebula} Albert, J., Aliu, E., Anderhub, H., et al. (MAGIC Collaboration) 2008, \APJ, 674, 1037
%\bibitem[Albert et al.(2009)]{magic_LSI61} Albert, J., Aliu, E., Anderhub, H., et al. (MAGIC Collaboration) 2009, \APJ, 693, 303
%\bibitem[Aliu et al.(2009)]{veritas_psr_limit_HDGS} Aliu, E., for the VERITAS Collaboration, 2009, ``Search for VHE $\gamma$-ray emission in the vicinity of selected pulsars of the Northern Sky with VERITAS'', AIP Conference Proceedings, 1085, 324
%\bibitem[Aliu et al.(2008)]{magic_crab_psr} Aliu, E., Anderhub, H., Antonelli, L. A., et al. (MAGIC Collaboration) 2008, Science, 322, 1221
%\bibitem[Atwood et al.(2009)]{lat_technical} Atwood, W.~B., Abdo, A.~A., Ackermann, M., et al. (Fermi/LAT Collaboration) 2009, \APJ, 697, 1071
%\bibitem[Berge et al.(2007)]{berge07} Berge, D., Funk, S., \& Hinton, J. A. 2007, \AAP, 466, 1219
%\bibitem[Bertsch et al.(2000)]{Bertsch00} Bertsch, D.~L., Hartman, R. C., Hunter, S. D., et al.\ 2000, American Institute of Physics Conference Series, 510, 504
%\bibitem[Celik et al.(2009)]{lat_geminga} Celik, O., on behalf of the Fermi-LAT Collaboration, 2009, ``Fermi-LAT observations of the Geminga pulsar'', to be published in the proceedings of the 31st International Cosmic Ray Conference
\bibitem[Chaves et al.(2009a)]{hess_survey08} Chaves, R.~C.~G., Renaud, M., Lemoine-Goumard, M., \& Goret, P., for the H.E.S.S. Collaboration, 2009a, AIP Conference Proceedings \textbf{1085}, 219
\bibitem[Chaves et al.(2009b)]{hess_survey09} Chaves, R.~C.~G., on behalf of the H.E.S.S. Collaboration, 2009b, ``Extending the H.E.S.S. Galactic Plane Survey'', to be published in the proceedings of the 31st International Cosmic Ray Conference
%\bibitem[Chaves et al.(2009c)]{hess_1848_HDGS} Chaves, R.~C.~G., de O{\~n}a Wilhemi, E., \& Hoppe, S., for the H.E.S.S. Collaboration,  2009c, AIP Conference Proceedings, 1085, 372
%\bibitem[Cohen-Tanugi et al.(2009)]{LAT_GC_icrc09} Cohen-Tanugi, J., Pohl, M., Tibolla, O., Parent, D., Nuss, E. for the Fermi/LAT Collaboration, 2009, to be published in the proceedings of the 31st International Cosmic Ray Conference
%\bibitem[Djannati-Atai et al.(2008a)]{icrc07_1908+063} Djannati-Atai, A., de Jager, O.~C., Terrier, R., Gallant, Y.~A., Hoppe, S., for the H.E.S.S. Collaboration,\ 2008a, Proceedings of the 30th International Cosmic Ray Conference; Rogelio Caballero, Juan Carlos D'Olivo, Gustavo Medina-Tanco, Lukas Nellen, Federico A. S\'anchez, Jos\'e F. Vald\'es-Galicia (eds.); Universidad Nacional Aut\'onoma de M\'exico, Mexico City, Mexico, 2, 823
%\bibitem[Djannati-Atai et al.(2008b)]{icrc07_1833_105} Djannati-Atai, A., O\~na-Wilhelmi, E., Renaud, M., Hoppe, S., for the H.E.S.S. Collaboration,\ 2008b, Proceedings of the 30th International Cosmic Ray Conference; Rogelio Caballero, Juan Carlos D'Olivo, Gustavo Medina-Tanco, Lukas Nellen, Federico A. S\'anchez, Jos\'e F. Vald\'es-Galicia (eds.); Universidad Nacional Aut\'onoma de M\'exico, Mexico City, Mexico, 2, 863
%\bibitem[Dubois et al.(2009)]{hess_velaX_2009} Dubois, F., Gl\"uck, B., de Jager, O. C., et al. 2009, Proceedings of the 31st ICRC, {\L}\'od\'z, Poland
%\bibitem[van Eldik et al.(2008)]{hess_GC_pos_icrc2007} van Eldik, C., Bolz, O., Braun, I., Hermann, G., Hinton, J. \& Hofmann, W.\ 2008, Proceedings of the 30th International Cosmic Ray Conference; Rogelio Caballero, Juan Carlos D'Olivo, Gustavo Medina-Tanco, Lukas Nellen, Federico A. S\'anchez, Jos\'e F. Vald\'es-Galicia (eds.); Universidad Nacional Aut\'onoma de M\'exico, Mexico City, Mexico, 2, 589
%\bibitem[Feldman \& Cousins(1998)]{feldman98} Feldman, G. J., \& Cousins, R. D. 1998, Phys. Rev. D., 57, 3873
%\bibitem[Fiasson et al.(2009)]{hess_w51} Fiasson, A., Marandon, V., Chaves, R.~C.~G., Tibolla, O., for the H.E.S.S. Collaboration, 2009, ``Discovery of a VHE gamma-ray source in the W51 region'', to be published in the proceedings of the 31st International Cosmic Ray Conference
%\bibitem[Finnegan et al.(2009)]{veritas_geminga} Finnegan, G., on behalf of the VERITAS Collaboration, 2009 ``Search for TeV emission from Geminga by VERITA'', to be published in the proceedings of the 31st International Cosmic Ray Conference
%\bibitem[Funk et al.(2004)]{funk04} Funk, S., Hermann, G., Hinton, J. A. et al.\ 2004, Astropart. Phys., 22, 285
\bibitem[Funk et al.(2008)]{funk08} Funk, S., Reimer, O., Torres, D. F., \& Hinton, J. A.\ 2008, \APJ 679 1299
%\bibitem[Gargano et al.(2009)]{lat_egr_psr} Gargano, F. for the Fermi/LAT Collaboration, 2009, ``Spectral analysis of EGRET pulsars'', to be published in the proceedings of the 31st International Cosmic Ray Conference
%\bibitem[Grondin et al.(2009)]{lat_crab} Grondin, M.-H., on behalf of the Fermi-LAT Collaboration, 2009, ``Fermi-LAT observations of the Crab nebula and pulsar'', to be published in the proceedings of the 31st International Cosmic Ray Conference
%\bibitem[Halpern et al.(2001)]{PSR2229_discovery_halpern01} Halpern, J.~P., Camilo, F., Gotthelf, E.~V., et al.\ 2001, \APJL, 552, L125
\bibitem[Hartman et al.(1999)]{3rd_egret_cat} Hartman, R.~C., Bertsch, D.~L., Bloom, S.~D., et al. 1999, \APJS 123 79
%\bibitem[Hillas(1996)]{hillas96} Hillas, A. M. 1996, Space~Sci.~Rev., 75, 17
%\bibitem[Hoppe et al.(2008)]{hess_survey_2007} Hoppe, S., for the H.E.S.S. Collaboration,\ 2008, Proceedings of the 30th International Cosmic Ray Conference; Rogelio Caballero, Juan Carlos D'Olivo, Gustavo Medina-Tanco, Lukas Nellen, Federico A. S\'anchez, Jos\'e F. Vald\'es-Galicia (eds.); Universidad Nacional Aut\'onoma de M\'exico, Mexico City, Mexico, 2, 579, preprint[arXiv:0710.3528]
%\bibitem[Hoppe et al.(2009)]{hess_1706} Hoppe, S., de O{\~n}a Wilhemi, E., Kh{\'e}lifi, B., Chaves, R.~C.~G., de Jager, O.~C., Stegmann, C., Terrier, R., for the H.E.S.S.~Collaboration, 2009, ``Detection of very-high-energy gamma-ray emission from the vicinity of PSR B1706-44 with H.E.S.S.'', to be published in the proceedings of the 31st International Cosmic Ray Conference, preprint[arXiv:0906.5574]
%\bibitem[Hinton(2009)]{Hinton09_review} Hinton, J.\ 2009, New Journal of Physics, 11, 055005
%\bibitem[de Jager et al.(1996)]{egret_crab_nebula} de Jager, O.~C., Harding, A.~K., Michelson, P.~F., Nel, H.~I., Nolan, P.~L., Sreekumar, P., \& Thompson, D.~J.\ 1996, \APJ, 457, 253
%\bibitem[Lamb \& Macomb(1997)]{GeV_cat} Lamb, R.~C., \& Macomb, D.~J.\ 1997, \APJ, 488, 872
%\bibitem[Lemoine-Goumard  et al.(2009)]{lat_velaX} Lemoine-Goumard, M., and Grondin, M.-H., on behalf of the Fermi-LAT Collaboration and the Pulsar Timing Consortium, 2009, ``Fermi-LAT observations of the Vela X region'', to be published in the proceedings of the 31st International Cosmic Ray Conference
%\bibitem[Li \& Ma(1983)]{LiMa83} Li, T.-P., \& Ma, Y.-Q.\ 1983, \APJ, 272, 317
%\bibitem[Mayer-Hasselwander et al.(1994)]{egret_geminga} Mayer-Hasselwander, H.~A., Bertsch, D.~L., Brazier, K.~T.~S., et al.\ 1994, \APJ, 421, 276
%\bibitem[Ng et al.(2005)]{ng_chandra_rabbit} Ng, C.-Y., Roberts, M.~S.~E., \& Romani, R.~W.\ 2005, \APJ, 627, 904
%\bibitem[Ohm et al.(2009)]{hess_westerlund1_1648} Ohm, S., Horns, D., de O{\~n}a Wilhemi, E., for the H.E.S.S. Collaboration, 2009, ``H.E.S.S. observations towards the massive stellar cluster Westerlund 1'', to be published in the proceedings of the 31st International Cosmic Ray Conference
%\bibitem[Reimer et al.(2008)]{reimar_icrc2007} Reimer, O., Funk, S., Torres, D.~F., \& Hinton, J.\ 2008, Proceedings of the 30th International Cosmic Ray Conference; Rogelio Caballero, Juan Carlos D'Olivo, Gustavo Medina-Tanco, Lukas Nellen, Federico A. S\'anchez, Jos\'e F. Vald\'es-Galicia (eds.); Universidad Nacional Aut\'onoma de M\'exico, Mexico City, Mexico, 2, 613
%\bibitem[de los Reyes et al.(2009)]{magic_psr_limits} de los Reyes, R., Bednarek, W., Camara, M., \& M. Lopez for the MAGIC Collaboration, 2009, ``Upper limits for pulsars with MAGIC (2005/2006 observations)'', to be published in the proceedings of the 31st International Cosmic Ray Conference
%\bibitem[Thompson et al.(1996)]{egret_psr1706} Thompson, D.~J., Bailes, M., Bertsch, D.~L., et al.\ 1996, \APJ, 465, 385
%\bibitem[Tibolla et al.(2009a)]{HowCanFermiHelp_SciNeGHe08} Tibolla, O., on behalf of the H.E.S.S. Collaboration, 2009a, AIP Conference Proceedings, 1112, 211
%\bibitem[Tibolla et al.(2009b)]{hess_1741_HDGS} Tibolla, O., Komin, N., Kosack, K., \& Naumann-Godo, M., on behalf of the H.E.S.S. Collaboration, 2009b, ``A new source discovered close to the Galactic Center: HESS~J1741-302'', AIP Conference Proceedings, 1085, 249
%\bibitem[Zepka et al.(1996)]{PSR0631_discovery} Zepka, A., Cordes, J.~M., Wasserman, I., \& Lundgren, S.~C.\ 1996, \APJ, 456, 305



% IT IS VERY IMPORTANT TO LEAVE THE EMPTY LINE ABOVE AT THE END OF YOUR
% REFERENCE LIST !!!
\end{thebibliography}
\end{document}